\documentclass[12pt,amsmath,amssymb,a4paper,final]{iopart}

\usepackage{braket}
\usepackage{hyperref}
\usepackage{color,graphicx}
\usepackage{dcolumn}
\usepackage{bm}
\usepackage{amsfonts}
 \expandafter\let\csname equation*\endcsname\relax
   \expandafter\let\csname endequation*\endcsname\relax
\usepackage{amsmath}
\usepackage{iopams}
\usepackage{amssymb}
\usepackage{textcomp}
\usepackage{cite}

\usepackage{wasysym}
\usepackage{booktabs}

\begin{document}

\title[Orbital breathing in the computation of XA and RIXS in solids by WF methods]%
{Orbital breathing effects in the computation of x-ray {\em d}-ion spectra in solids by {\em ab
    initio} wave-function-based methods}

\author{Nikolay A~Bogdanov$\bm{^{1,2}}$, Valentina Bisogni$\bm{^{3,4}}$, Roberto Kraus$\bm{^5}$, Claude
  Monney$\bm{^{3,6}}$, Kejin Zhou$\bm{^{3,7}}$, Thorsten Schmitt$\bm{^3}$, Jochen Geck$\bm{^{5,8}}$, Alexander
  O~Mitrushchenkov$\bm{^9}$, Hermann Stoll$\bm{^{10}}$, Jeroen van den Brink$\bm{^1}$ and Liviu Hozoi$\bm{^1}$}

\address{$^1$Institute for Theoretical Solid State Physics, IFW Dresden, Helmholtzstr.~20, 01069
  Dresden, Germany} %
\address{$^2$Max Planck Institute for Solid State Research, Heisenbergstr.~1, 70569 Stuttgart,
  Germany} %
\address{$^3$Swiss Light Source, Paul Scherrer Insitute, CH-5232 Villigen PSI, Switzerland} %
\address{$^4$National Synchrotron Light Source II, Brookhaven National Laboratory, Upton, New York
  11973-5000, USA} %
\address{$^5$Institute for Solid State Research, IFW Dresden, Helmholtzstr.~20, 01069 Dresden,
  Germany} %
\address{$^6$Department of Physics, Universit\"at Z\"urich, Winterthurerstrasse 190, 8057 Z\"urich,
  Switzerland} %
\address{$^7$Diamond Light Source, Harwell Science \& Innovation Campus, Didcot, Oxfordshire OX11
  0DE, UK} %
\address{$^8$Chemistry and Physics of Materials, Paris Lodron University of Salzburg, Hellbrunner
  Str.~34, 5020 Salzburg, Austria} %
\address{$^9$Laboratoire de Mod\'elisation et Simulation Multi Echelle, MSME UMR 8208 CNRS,
  Universit\'e Paris-Est, Marne-la-Vall\'ee, France} %
\address{$^{10}$Institut f\"ur Theoretische Chemie, Universit\"at Stuttgart, Pfaffenwaldring 55, 70569
  Stuttgart, Germany} %

\ead{n.bogdanov@fkf.mpg.de, l.hozoi@ifw-dresden.de}

\date{12 October 2016}


\begin{abstract}
In existing theoretical approaches to core-level excitations of transition-metal ions in solids
relaxation and polarization effects due to the inner core hole are often ignored or described
phenomenologically.
Here we set up an {\it ab initio} computational scheme that explicitly accounts for such
physics in the calculation of x-ray absorption and resonant inelastic x-ray scattering spectra.
Good agreement is found with experimental transition-metal $L$-edge data for the strongly
correlated $d^9$ cuprate Li$_2$CuO$_2$, for which we determine the absolute scattering
intensities.
The newly developed methodology opens the way for the investigation of even more complex
$d^n$ electronic structures of group VI\,B to VIII\,B correlated oxide compounds. 
\end{abstract}


\maketitle

\section{Introduction}  
Some of the most interesting properties of a molecule or a solid concern the response to an
external electromagnetic field.
This response also provides clues on the actual electronic structure of the system and is
therefore a main aspect that is analyzed in experimental labs. 
For high-energy incoming photons as used in x-ray absorption or photoemission measurements,
a challenging task on the theoretical and computational side is a reliable description
of the changes that occur in the electronic environment after creating a localized electron
vacancy, be it in a core or valence-shell level.
Such changes are also referred to as charge relaxation, orbital breathing, or screening effects
and can be classified into intra-atomic and extra-atomic contributions \cite{book_fulde_12,
book_veenendal_15}.
Here we outline an {\it ab initio} wave-function-based methodology that allows to explicitly
describe the readjustment of the charge distribution in the `vicinity' of an excited electron
by performing separate self-consistent-field (SCF) optimizations for the different electron
configurations.
While this idea of using individually optimized wave functions has been earlier employed for
the interpretation of x-ray absorption (XA) \cite{XA_noci_yang96,XA_raspt2_pinjari14} and
resonant inelastic x-ray scattering (RIXS) \cite{rixs_raspt2_kunnus16,rixs_raspt2_guo16}
spectra of small organic molecules \cite{XA_noci_yang96} and of Fe-based complexes in solution
\cite{XA_raspt2_pinjari14,rixs_raspt2_kunnus16,rixs_raspt2_guo16}, we here apply it to the
calculation of XA and RIXS excitations and cross sections of a transition-metal ion in a
solid-state matrix and compare the computational results to fresh experimental data.
Li$_2$CuO$_2$, a strongly correlated $d$-electron system, is chosen as a test case.
We find large SCF relaxation effects of $\gtrsim$10\,eV for the Cu core-hole excited states.
Using such SCF many-body wave functions, all trends found experimentally for the incoming-photon
incident-angle and polarization dependence are faithfully reproduced in the computed RIXS
spectra.
This also allows us to determine RIXS intensities on an absolute scale, from which we
compute a resonant scattering enhancement of the order of $10^5$.

%
RIXS is a powerful and fast-developing technique to measure a diversity of different 
elementary excitations in correlated electron systems \cite{RIXS_RevModPhys_11,book_veenendal_15},
for example, dispersive magnetic modes \cite{rixs_mgn_2009}, orbital excitations
\cite{rixs_d1_2009,Moretti_CFfits_2011,rixs_orbiton_2012} and phonons \cite{rixs_phonons_2013,rixs_phonons_2016}.
As the interaction between x-ray photons and matter is relatively strong --- much stronger than
for instance the interaction of neutrons with matter --- these modes can be measured on very
small sample volumes, for example, cuprate nanostructures that are only a few unit cells thick
\cite{rixs_nano_mark_2012,rixs_nano_lucio_2012}.
Also, the scattered x-ray photon leaves the material under study behind with a low-energy 
elementary excitation that may be directly compared to the elementary response of other
inelastic scattering techniques.
One complication in RIXS is however posed by the resonant character of the technique, which
implies the presence of a resonant intermediate state.
The intermediate state is transient but strongly perturbed with respect to the ground state due
to the absorption of a high-energy x-ray photon which results in the creation of a hole in the
electronic core.
The difficulties in the calculation and interpretation of RIXS spectra arise from the fact
that the intermediary electron configuration determines all the transition probabilities from
the initial to the final state and as such affects the scattering cross sections of the
different excitations. 

Attempts to compute full RIXS spectra for correlated, $d$-electron open-shell systems have been
made with both model-Hamiltonian approaches
\cite{book_veenendal_15,RIXS_RevModPhys_11,rixs_ament_2007,rixs_UHB_freericks_2012,rixs_UHB_igarashi_2013,rixs_UHB_tohyama_2015,rixs_cdw_benjamin_2015,rixs_NiO_wray_2012,rixs_tanaka_2011,rixs_haverkort_2012}
and {\it ab initio} wave-function-based quantum chemistry methods 
\cite{rixs_qc_josefsson_2012,rixs_qc_kunnus_2013,rixs_qc_maganas_2014,rixs_raspt2_kunnus16,rixs_raspt2_guo16},
to address either the $d$-shell multiplet structure
\cite{rixs_NiO_wray_2012,rixs_tanaka_2011,rixs_haverkort_2012,rixs_qc_josefsson_2012,rixs_qc_kunnus_2013,rixs_raspt2_kunnus16,rixs_raspt2_guo16}
or Mott-Hubbard physics 
\cite{rixs_ament_2007,rixs_UHB_freericks_2012,rixs_UHB_igarashi_2013,rixs_UHB_tohyama_2015,rixs_cdw_benjamin_2015}.
Yet, in the solid-state context, one important aspect that is either missing or described just
phenomenologically in earlier computational work is the valence-shell charge relaxation in 
response to the creation of a core hole in the intermediate RIXS state.
To treat such effects in a reliable way, we choose to carry out independent SCF optimizations
for the many-body wave functions describing the reference $d^n$ and core-hole $c^{\star}d^{n+1}$
configurations.
This is achieved through multiconfiguration SCF (MCSCF) calculations \cite{book_QC_00} on
relatively large but nevertheless finite clusters with appropriate solid-state embedding.
The individual MCSCF optimizations obviously lead to sets of nonorthogonal orbitals.
Matrix elements (ME's) between wave functions expressed in terms of nonorthogonal orbitals can
be however computed with dedicated quantum chemistry algorithms
\cite{noci_broer_81,SI_malmqvist_86,SI_mitrushchenkov_07,XA_noci_yang96}.
Not only is our approach unbiased this way with regard to strong charge readjustment effects,
it also allows to calculate absolute RIXS cross sections --- core information for any scattering
technique.
Our results for the latter can directly be tested experimentally.

\section{Cu {\em L}-edge excitations in Li$_\textbf{2}$CuO$_\textbf{2}$}
\begin{figure}[!b]
\begin{indented}
\item[] \includegraphics[width=.6\linewidth]{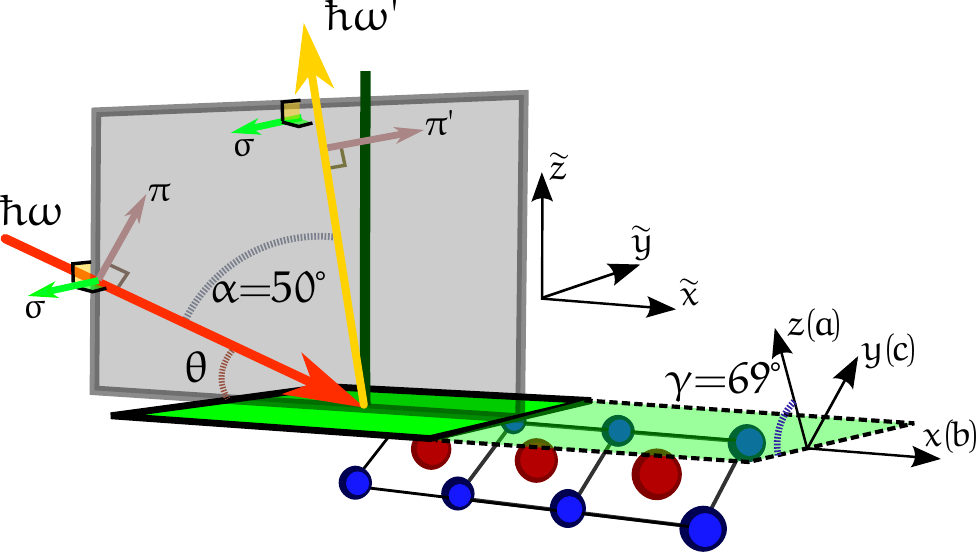}
\caption{
Sketch of the scattering geometry in Li$_2$CuO$_2$, see text.
The small spheres depict a sequence of three edge-sharing CuO$_4$ plaquettes.
}
\label{fig:RIXS_geom}
\end{indented}
\end{figure}

Li$_2$CuO$_2$ is being extensively investigated in the context of low-dimensional quantum magnetism.
We here address the Cu 2$p$ to 3$d$ ($L$-edge) excitations of this compound.
Quantum chemistry calculations were carried out on a [Cu$_3$O$_8$Li$_{16}$] cluster consisting
of one reference CuO$_4$ plaquette for which Cu 2$p$--3$d$ and 3$d$--3$d$ excitations are explicitly
computed, two other nearest-neighbor (NN) CuO$_4$ plaquettes (see \autoref{fig:RIXS_geom})
and 16 adjacent Li ions.
The surrounding solid-state matrix was modeled by an array of point charges optimized to reproduce
the ionic Madelung field. 
To reduce the computational effort and the complexity of the subsequent analysis, the NN
Cu$^{2+}$ $S\!=\!1/2$ 3$d^9$ ions were represented by closed-shell Zn$^{2+}$ 3$d^{10}$ species.
Scalar relativistic effects were taken into account using the Douglas-Kroll-Hess methodology
\cite{DKH_1974,DKH_1986}.
We used all-electron relativistic basis sets of valence-shell quadruple-zeta
quality with polarization functions for Cu and O sites of the central plaquette along with valence
double-zeta basis functions for the neighboring Cu, O and Li ions \cite{Balabanov_2005,Dunning_1989,Prascher_2010}.
All computations were performed with the {\sc molpro} quantum chemistry package \cite{Molpro12}.

The scattering process involves absorption of a single photon with known energy, momentum and
polarization $\{\hbar\omega,\vec{k},\varepsilon\}$ and emission of another photon characterized
by the parameters $\{\hbar\omega',\vec{k}',\varepsilon'\}$.
The RIXS double differential cross section is \cite{book_veenendal_15,book_Dirac_TPQM}\,:
\begin{equation}
\begin{aligned}[l]
{ d^2\sigma^{\mathrm{RIXS}}_{\vec{k},\varepsilon} \over d\Omega'd\hbar\omega'}\!=\!
{\alpha^2 \hbar^2 \over c^2} \omega \omega'^3
&
        \sum_{j}{{1}\over{g_{\mathrm{gs}}}}
         \sum_{k}
\!\left |\sum_{l}{\bra{\Psi_{\mathrm{fs}}^k}\vec{\epsilon}\,'\!\cdot\! \vec{R} \ket{\Psi_{\mathrm{c^{\star}}}^l}
                  \bra{\Psi_{\mathrm{c^{\star}}}^l}\vec{\epsilon} \!\cdot\! \vec{R} \ket{\Psi_{\mathrm{gs}}^j}
                  \over  {E_{\mathrm{gs}}^j + \hbar\omega - E_{\mathrm{c^{\star}}}^l + i\Gamma_{\mathrm{c^{\star}}}^l}/2}
 \right |^2& \\
&\times
 {{\Gamma_{\mathrm{fs}}^k/ 2\pi} \over
 \left( E_{\mathrm{gs}}^j + \hbar\omega - E_{\mathrm{fs}}^k - \hbar\omega' \right)^2 + \left( \Gamma_{\mathrm{fs}}^k\right)^2/4} \ ,
\end{aligned}
\label{RIXScs}
\end{equation}
where $\alpha$ is the fine structure constant, $\vec{\epsilon}$ and $\vec{\epsilon}\,'$ are
polarization vectors of the incoming and outgoing photons and $\vec{R}$ is a site specific
position operator.
The lifetimes for the excited core-hole and final states are $\Gamma_{\mathrm{c^{\star}}}$ and
$\Gamma_{\mathrm{fs}}$, respectively.
The summations take into account all available intermediate and final states and the possible
degeneracy of the ground state, $g_\mathrm{gs}$.

Many-body wave functions, eigenvalues and dipole transition ME's were computed at both the 
MCSCF and multireference configuration-interaction (MRCI) levels of theory \cite{book_QC_00}.
The active orbital space includes all five Cu 3$d$ orbitals for the Cu 3$d^9$ states and
five Cu 3$d$ orbitals plus the Cu 2$p$ functions for the intermediate 2$p^5$3$d^{10}$ states.
In order to prevent for the latter MCSCF states 2$p^5$3$d^{10}$$\rightarrow$2$p^6$3$d^9$ `de-excitation',
the orbital optimization was carried out in two steps in the MCSCF calculations, following a
freeze-and-thaw orbital relaxation (FTOR) scheme\,:
(i)  optimize first the core-level Cu 2$p$ orbitals, while freezing all other orbitals;
(ii) freeze next the Cu 2$p$ functions and optimize the rest.
Repeating these two steps until there is no further change in energy (averaged over all
2$p^5$3$d^{10}$ states) `keeps' the hole down in the core and allows the valence and semi-valence
shells to relax and polarize, both on-site and at neighboring sites.
Appropriate changes were further operated in the configuration-interaction module of {\sc molpro}
to preclude 2$p^5$3$d^{10}$$\rightarrow$2$p^6$3$d^9$ de-excitation in the MRCI wave functions.
The dipole transition ME's between the independently optimized $d^9$ and $c^{\star}d^{10}$ groups
of states were derived according to the procedure described in \cite{SI_mitrushchenkov_07}.
%
An approach in the same spirit, based on separate multiconfiguration SCF optimizations for the
different $d^n$ states but a subsequent treatment of further correlation effects by perturbational
methods, have been recently applied for the computation of XA and RIXS spectra of iron 
hexacyanides in solution \cite{XA_raspt2_pinjari14,rixs_raspt2_kunnus16,rixs_raspt2_guo16}.
In earlier work in the solid-state context, this idea of using individually optimized
wave functions has been employed for the interpretation of x-ray photoelectron spectra
\cite{noci_degraaf_99,noci_hozoi_06}, the computation of quasiparticle band structures
\cite{diamond_alex_14,mgo_hozoi_07,CuO_hozoi_07} and the analysis of superexchange virtual
excitations \cite{noci_vanoosten_96,noci_hozoi_03}.

The scattering geometry and details on how the polarization analysis is carried out are specified
in \autoref{fig:RIXS_geom}.
The unit cell of Li$_2$CuO$_2$ is defined in terms of skew axes, where the angle between the
$\left(100\right)$ and $\left(101\right)$ planes is $\gamma\!\approx\!\mathrm{69}^{\circ}$,
not $\mathrm{45}^{\circ}$ as in the cubic case.
The samples could only be cleaved perpendicular to the $\left[ 101\right]$ axis
\cite{Wizent_2014}.
The scattering plane thus incorporates the crystallographic $b$ axis and is perpendicular to 
the $\left ( 101 \right )$ plane.
The $b$ axis matches the direction of the chain of edge-sharing CuO$_4$ plaquettes, see
\autoref{fig:RIXS_geom}.
The `included' scattering angle (between incoming and outgoing light beams) was fixed to
$\alpha\!=\!180^{\circ}\!\!-\!130^{\circ}\!\!=\!\mathrm{50}^{\circ}$, while the incident 
angle $\theta$ (between the sample plane and the incoming photon direction) was varied from 
10$^{\circ}$ to 125$^{\circ}$.
The incoming light was linearly polarized, either perpendicular to the scattering plane
($\sigma$ polarization) or within the scattering plane ($\pi$ polarization).
Since no polarization analysis was performed for the outgoing radiation, we carry out for 
the latter an `incoherent' summation over the two independent polarization directions 
(the vector $\pi'$ is introduced to this end).

For the analysis of the quantum chemistry wave functions we use a local coordinate frame
$\{x,y,z\}$ having the $z$ axis perpendicular to the CuO$_4$ plaquette and $x$ along the chain.
The transformation to the setting used on the experimental side ($\{\tilde{x},\tilde{y},\tilde{z}\}$,
see \autoref{fig:RIXS_geom}) is however straightforward.
The rotation of the $\sigma$, $\pi$ and $\pi'$ vectors as function of the angle $\theta$ is
also described by simple geometrical relations\,:
\begin{equation}
  \begin{aligned}
    \vec{D}_{\sigma}&=\vec{D}_{\tilde{y}} \ ,&\\
    \vec{D}_{\pi}&=\vec{D}_{\tilde{x}} \sin{\theta}+\vec{D}_{\tilde{z}}
    \cos{\theta} \ ,&\\
    \vec{D}_{\pi'}&=\vec{D}_{\tilde{x}}
    \sin{(\theta+\alpha)}+\vec{D}_{\tilde{z}} \cos{(\theta+\alpha)} \ ,&
  \end{aligned}
  \label{D_pol}
\end{equation}
where 
$\vec{D}^{ij}\!=\!\bra{\Psi_{\mathrm{c^{\star}}}^i}e\cdot\vec{R}\ket{\Psi_{d^9}^j}$ stands for dipole
transition ME's, with superscripts dropped in (2).
By summing up over the outgoing polarization directions, (1) changes to
\begin{equation}
\begin{aligned}[l]
I(\hbar \omega,\hbar \omega', \varepsilon, \theta)\!=\!
{ d^2\sigma^\mathrm{RIXS}_{\vec{k},\varepsilon} \over d\Omega'd\hbar\omega'}\!=\!
{\alpha^2 \hbar^2 \over e^4 c^2} \omega \omega'^3
&        \sum_{\varepsilon'} \sum_{\mathrm{gs}}{{1}\over{g_{\mathrm{gs}}}}
 \sum_{\mathrm{fs}}
\!\left | \sum_{\mathrm{c^{\star}}} {\bra{\mathrm{fs}} \vec{D}_{\varepsilon'}  \ket{\mathrm{c^{\star}}}
                 \bra{\mathrm{c^{\star}}} \vec{D}_{\varepsilon} \ket{\mathrm{gs}}
                 \over {E_{\mathrm{gs}}+\hbar\omega-E_{\mathrm{c^{\star}}}+i\Gamma_{\mathrm{c^{\star}}}}/2}
 \right |^2& \\
&\times
 {{\Gamma_{\mathrm{fs}}/ 2\pi} \over
 \left( E_{\mathrm{gs}}+\hbar\omega-E_{\mathrm{fs}}-\hbar\omega' \right)^2\!+\!\Gamma_{\mathrm{fs}}^2/4}.&
\end{aligned}
\label{RIXSintens}
\end{equation}
The natural widths $\Gamma_{\mathrm{c^{\star}}}$ and $\Gamma_{\mathrm{\mathrm{fs}}}$ are here set to
1 and 0.1 eV, respectively, typical values for Cu $L$-edge {RIXS}
\cite{RIXS_RevModPhys_11,Moretti_CFfits_2011}.

\begin{table}[!t]
\caption{
Relative energies (eV) for Cu$^{2+}$ 3$d^9$ and 2$p^5$3$d^{10}$ states in Li$_2$CuO$_2$.
The MRCI $d$-level splittings include Davidson corrections \cite{book_QC_00}.
The $x$ axis is taken along the chain of CuO$_4$ plaquettes and $z$ perpendicular to the
plaquette plane, see \autoref{fig:RIXS_geom}.
Each theoretical value stands for a spin-orbit Kramers doublet.
}
\label{tab:dd_split}
\begin{indented}
\lineup
\item[]\begin{tabular}{@{}llll}
\br
Hole orbital        &MCSCF+SOC  &MRCI+SOC   &Experiment\\
\ns
\mr
$d_{xy}$            &0          &0          &0  \\
$d_{x^2-y^2}$       &1.24       &1.55       &1.7\\
$d_{zx}$            &1.73       &2.02       &2.1\\
$d_{yz}$            &1.80       &2.09       &2.1\\
$d_{3z^2-r^2}$           &2.54       &2.85       &2.6\\[0.15cm]
$p_{3/2}$           &933.6; 933.8$^{\rm a}$
                                &932.5; 932.6
                                            &931.6\\
$p_{1/2}$           &954.4$^{\rm a}$
                                &953.3      &---  \\
\br
\end{tabular}
\item[] $^{\rm a}$ {944.9, 945.1 and 965.2 eV by using orbitals optimized for the 3$d^9$
    configuration.} 
\end{indented}
\end{table}

The RIXS experiments were performed at the {\sc adress} beamline of the Swiss Light Source (Paul Scherrer
Institute) using the {\sc saxes} spectrometer \cite{rixs_adress_2010}.
The RIXS spectra were recorded using a scattering angle of 130$^{\circ}$.
The combined energy resolution at the Cu $L$-edge was 130 meV.
Given its hygroscopic character, the crystal was cleaved in-situ at $T\!=\!20$ K, as described
in \cite{rixs_ZR_2013}.
The measurements were recorded at the same temperature, $T\!=\!20$ K.

\section{Results and discussion}

Computed excitation energies are compared to experimental data in \autoref{tab:dd_split}.
While XA and RIXS spectra have been earlier measured for Li$_2$CuO$_2$ at the O $K$-edge
\cite{Learmonth_2007,rixs_ZR_2013}, Cu $L$-edge data have not been reported so far on this
material.
The Cu 2$p$ core-hole states are strongly affected by spin-orbit couplings (SOC's): the 2$p^5$
$j\!=\!\mathrm{3}/\mathrm{2}$ quartet and $j\!=\!\mathrm{1}/\mathrm{2}$ doublet are separated
by $\approx$20\,eV.
The effect of SOC is on the other hand fairly small for the Cu 3$d^9$ valence states and brings 
only tiny corrections to those relative energies.
The $p_{3/2}$ $p^6d^9$\,$\rightarrow$\,$p^5d^{10}$ excitation energy as computed at the MCSCF
level overestimates by about 2 eV the experimental result.
But without a separate SCF optimization of the 2$p^5$3$d^{10}$ states and using orbitals optimized
for the ground-state 2$p^6$3$d^9$ configuration, as in previous $L$-edge electronic-structure calculations
\cite{rixs_qc_carniato_2009,rixs_qc_kavcic_2010,rixs_qc_josefsson_2012,rixs_qc_kunnus_2013,rixs_qc_maganas_2014},
this energy difference is larger by a factor of 5, i.e., about 11 eV (see footnote in
\autoref{tab:dd_split}).
This strong effect that we quantify here has to do with readjustment of the electronic charge
within the Cu second and third atomic shells upon the 2$p$\,$\rightarrow$\,3$d$ transition and
further to orbital polarization at neighboring O sites.
It is even stronger than the short-range relaxation and polarization effects for $(N\!\pm\!1)$
processes (i.e., complete removal of a $\pm e$ charge) within the valence levels of, e.g., MgO
or diamond \cite{mgo_hozoi_07,diamond_alex_14}.
Obviously, such strong charge readjustment can in principle significantly affect the dipole
transition ME's employed for deriving the RIXS intensities.

\begin{figure}[bt]
  \begin{indented}
\item[]    \includegraphics[width=0.8\linewidth]{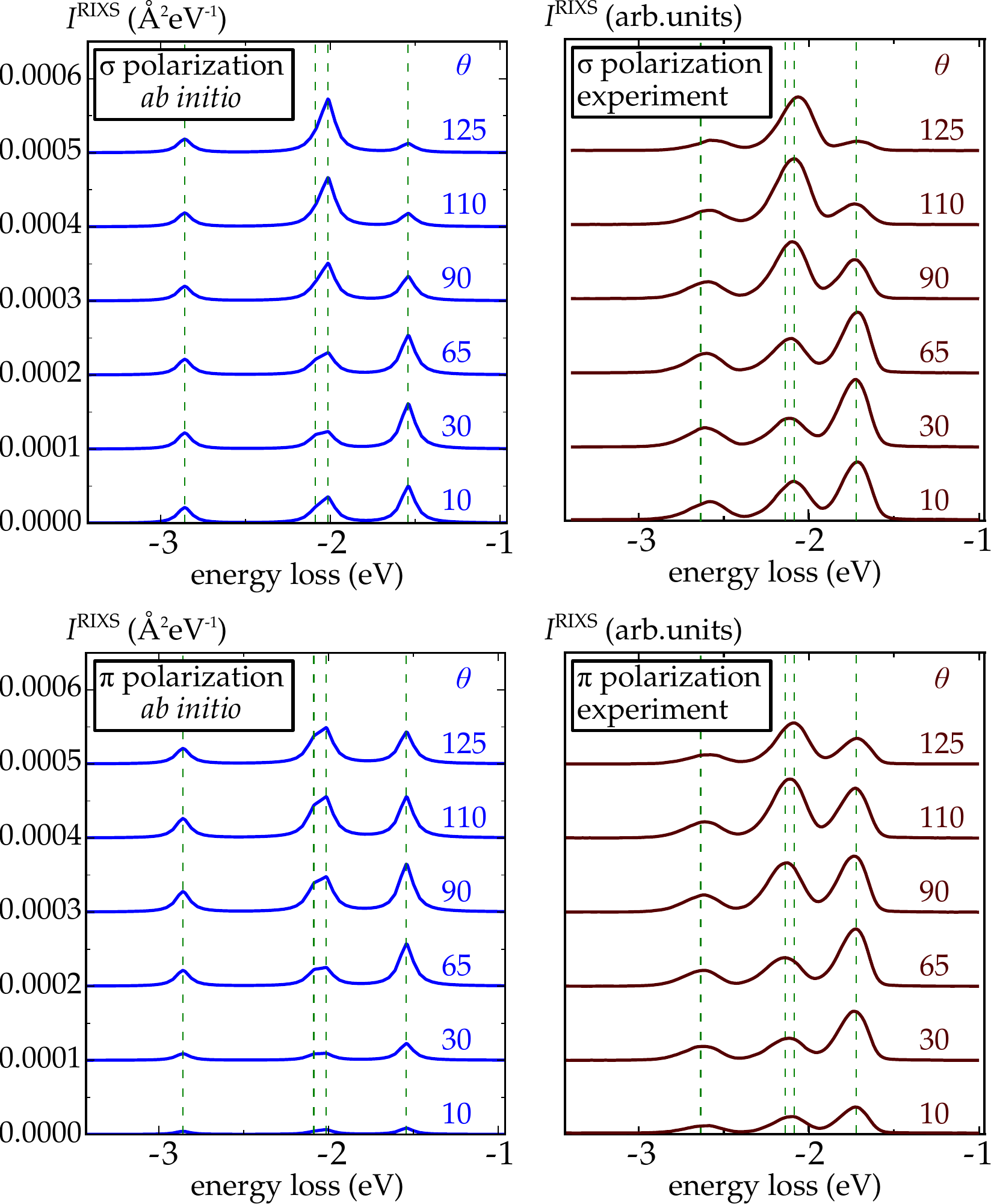}
    \caption{Calculated (MRCI+SOC) versus experimental $L_3$-edge double differential cross
      sections as function of the incident angle $\theta$.}
    \label{fig:RIXS_pol_L3}
  \end{indented}
\end{figure}

As seen in \autoref{tab:dd_split}, the MRCI treatment yields additional corrections of
$\gtrsim$\,1 eV to the $p_{3/2}$ $p^6d^9$\,$\rightarrow$\,$p^5d^{10}$ excitation energies, which
brings the {\it ab initio} values within 1 eV of the measured $p_{3/2}$ XA edge.
It is worth pointing out that longer-range polarization effects are not very important for 
relatively localized excitations that conserve the number of electrons in the system such as the
$L$-edge $2p$\,$\rightarrow$\,$3d$ transitions.
This is one feature that makes the accurate modeling (and the interpretation) of $L$-edge $d$-metal
RIXS spectra easier than of, e.g., photoemission spectra \cite{book_fulde_12,QPs_hirsch_2014,QPs_reining_2015,xps_rehr_2015},
once a reliable many-body scheme is set up for describing the `local' multiplet structure.

\begin{figure}[t]
  \label{fig:RIXS_plane}
\begin{indented}
\item[]  \includegraphics[width=0.95\linewidth]{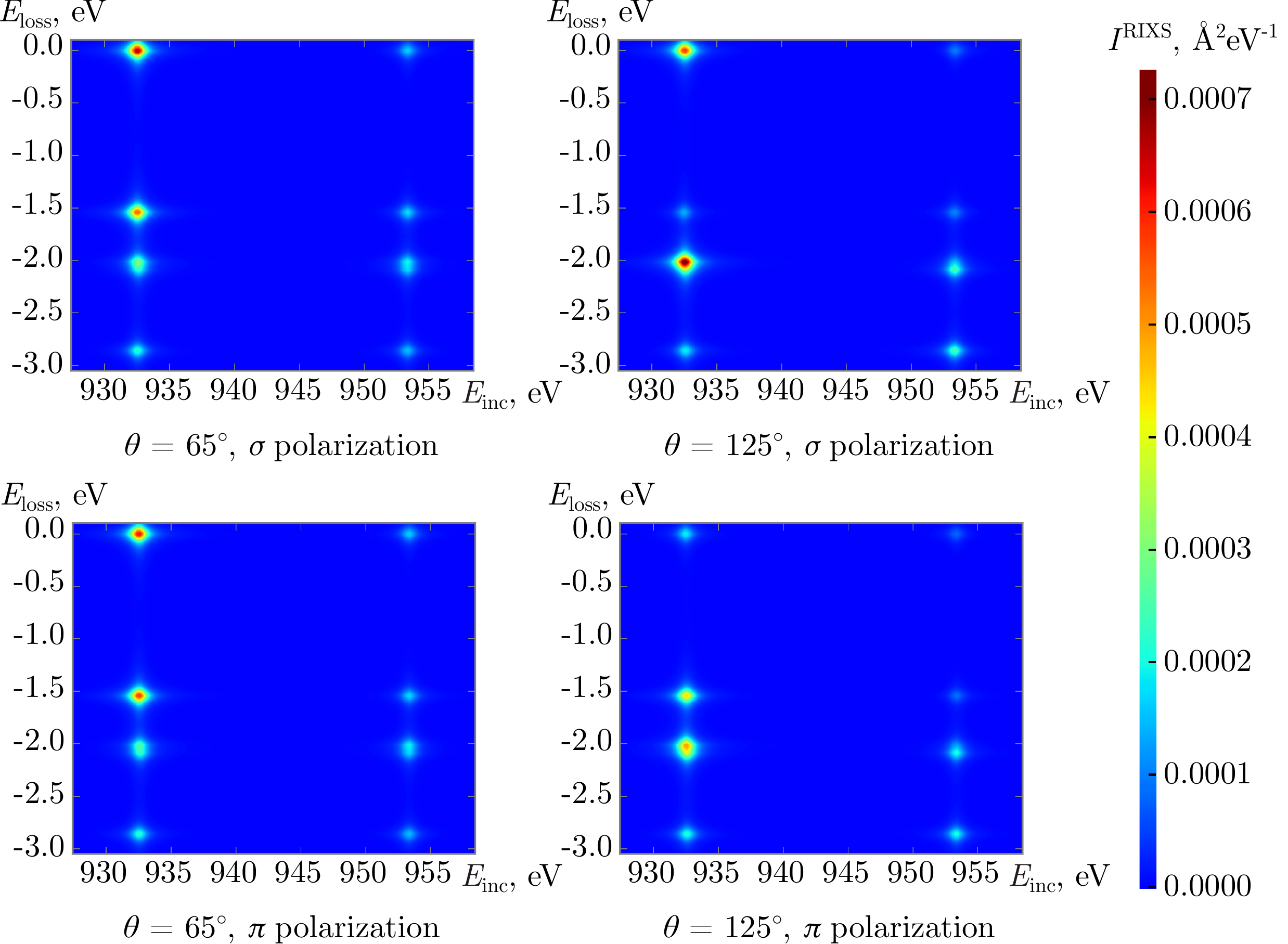}
  \caption{
Theoretical $I(E_{\mathrm{inc}},E_{\mathrm{loss}})$ RIXS plots for Li$_2$CuO$_2$ using the
geometrical setup employed in the measurements.
In addition to the Cu $L_3$-edge excitations, the $L_2$ `signal' is also visible.
}
\end{indented}
\end{figure}

{\it Ab initio} results for full Cu $L_3$-edge RIXS spectra are shown in \autoref{fig:RIXS_pol_L3}
along with experimental data.
The theoretical spectra are obtained on an absolute intensity scale, an aspect on which we
shall elaborate in the following.
The evolution of the RIXS intensities when modifying the incident angle $\theta$ is different
for each of the main peaks --- see also the $I(E_{\mathrm{inc}},E_{\mathrm{loss}})$ plots in
\autoref{fig:RIXS_plane} --- and this specific dependence can be used to unmistakably identify
the origin of each particular feature.
The quantum chemistry data agree well with the experiment, with all major trends nicely
reproduced:
the intensities of the two peaks in the range of 2--3 eV, in particular, clearly display distinct
behavior with increasing incident angle --- while the higher-energy excitation loses spectral
weight, the one at about 2 eV gradually acquires more and more intensity.
The calculations also reveal that the experimental feature at $\approx$2 eV has to do with two
different possible final states, with the 3$d$ hole in either the Cu $d_{yz}$ or $d_{zx}$ orbital.
Due to $x$--$y$ anisotropy, the $yz$ and $zx$ components are slightly split apart but not 
strongly enough to be resolved as two individual peaks in experiment.
The anisotropic environment splits as well the computed $p_{3/2}$ states (\autoref{tab:dd_split}),
into two sets of doublets.
With regard to peak positions, the agreement with the experimental data is within 0.15 eV
for the $d_{x^2-y^2}$, $d_{yz}$ and $d_{zx}$ hole states.
For the $d_{3z^2-r^2}$ hole state, the MRCI calculations overestimate the relative energy
by 0.25 eV, $\approx$10\%.

One achievement with the {\it ab initio} methodology is that absolute values for the dipole
transition ME's are obtained, not just relative estimates as in semi-empirical schemes.
In particular, it is possible to calculate absolute cross sections for the RIXS process using
\autoref{RIXSintens}.
The theoretical plots provided in \autoref{fig:RIXS_pol_L3} and \autoref{fig:RIXS_plane} are
based on such absolute values.
Having the latter, one can quantify specific `scattering properties' of the material, e.g.,
how many photons of particular energy and polarization will be scattered by a given atomic site
in a given direction.
No results have been reported so far for absolute experimental RIXS intensities, although such
measurements should in principle be possible.
Absolute intensities have been reported however for non-resonant inelastic x-ray scattering
(NIXS) \cite{Tischler_2003,Gurtubay_2005}.
Since the RIXS intensities are expected to be a few orders of magnitude larger than in NIXS 
\cite{RIXS_RevModPhys_11}, it is then instructive to compare our computational results with
the available NIXS data.
For direct comparison with the dynamical structure factor per volume unit measured by NIXS,
$s$,    
we must divide the cross sections from \autoref{RIXSintens} by the product of the unit cell
volume per Cu site, the squared classical electron radius and the ratio $\omega'/\omega\!\approx\!1$
and additionally account for the fact that in RIXS we integrate over all $\varepsilon'$.
By doing so, the initial value $I^{\mathrm{RIXS}}\!\sim\!\!10^{-4}$\,\AA$^{2}$eV$^{-1}$ 
translates into intensities $\sim\!\!10^3$\,\AA$^{-3}$eV$^{-1}$, about $10^{5}$ times larger
than typical values of $s$ in NIXS ($\sim\!10^{-2}$\,\AA$^{-3}$eV$^{-1}$) \cite{Gurtubay_2005}.

\section{Conclusions}
In sum, we introduce a multireference configuration-interaction-type of methodology for the 
{\it ab initio} calculation of transition-metal $L$-edge excitations in solids, i.e., both XA
energies and RIXS cross sections for the $2p\!\rightarrow\!3d\!\rightarrow\!2p$ transitions.
The results obtained with this computational scheme show good agreement with experimental data,
for both excitation energies and polarization- and angle-dependent intensities.
Orbital relaxation effects in the presence of the core hole are taken into account by performing two
independent MCSCF calculations and utilising afterwards a nonorthogonal configuration-interaction
type of approach for the computation of the dipole transition matrix elements
\cite{XA_noci_yang96,noci_broer_81,SI_malmqvist_86,SI_mitrushchenkov_07}.
The use of separately optimized orbitals brings an energy stabilization effect of 11\,eV for
the core-hole $c^{\star}d^{n+1}$ states as compared to the case where ground-state ($d^n$) orbitals
are employed.
While here we test this computational scheme for the more convenient case of a Cu$^{2+}$ 3$d^9$
ground state --- giving rise to a simple, closed-shell 3$d^{10}$ valence electron configuration
subsequent to the initial absorption process --- the procedure is applicable to other types of
correlated electron systems, $d^n$ or $f^n$.
Having the new methodology in hand, such investigations and more involved calculations for
arbitrary $n$ will be the topic of future studies.

\section{Acknowledgements}

This research has been partly funded by the German Science Foundation and the Swiss National
Science Foundation within the D-A-CH program (SNSF grant no.~200021L 141325 and DFG grant
GE 1647/3-1).
C.\,M.~acknowledges support by the Swiss National Science Fondation under grant no.~PZ00P2 154867.
L.\,H.~thanks the German Science Foundation for financial support (DFG grant HO 4427/2).
N.\,A.\,B.~and L.\,H.~acknowledge support from the High Performance Computing Center (ZIH) of the
Technical University Dresden (TUD) and thank H.\,Y.\,Huang and P.\,Fulde for fruitful discussions.

\section*{Appendix: Computational details of the quantum chemistry calculations}

Theoretical RIXS and XA excitation energies together with transition matrix elements of the dipole operator were
obtained by MCSCF and subsequent MRCI calculations \cite{book_QC_00}.
These computations were carried out on clusters which contain one central CuO$_4$ plaquette, the two
NN CuO$_4$ plaquettes and the nearby 16 Li ions.
The solid-state surroundings were modeled as a large array of point charges fitted to reproduce the
crystal Madelung field in the cluster region.
To obtain a clear picture on crystal-field effects at the central Cu site, we cut off the magnetic
couplings with the adjacent Cu ions by replacing the open-shell $d^9$ NN's with closed-shell
Zn$^{2+}$ $d^{10}$ species.
This is common procedure in quantum chemistry studies on transition-metal systems, see, e.g.,
Refs.\,\cite{qc_NNs_degraaf_99, Na2V2O5_hozoi_02, CuO2_dd_hozoi_11, NiO_domingo_12,Os227_bogdanov_12, SIA_Fe_maurice_2013}.

For all quantum chemistry calculations, the {\sc molpro} suite of {\it ab initio} programs
\cite{Molpro12} was employed.
We used all-electron relativistic basis sets from the standard {\sc molpro} library in conjunction
with the second-order Douglas-Kroll-Hess operator \cite{DKH_1974,DKH_1986}.
For the central Cu site, we utilized the $s$ and $p$ functions of the cc-pwCVQZ-DK basis set together with
$d$ and $f$ functions from the cc-pVQZ-DK compilation \cite{Balabanov_2005}.
For O ligands at the central CuO$_4$ plaquette, the $s$ and $p$ functions of the cc-pVQZ-DK basis set were used,
along with polarization $d$ functions of the cc-pVTZ-DK kit \cite{Dunning_1989}.
For other ions in the neighborhood of the central CuO$_4$ plaquette, valence double-zeta basis
functions of cc-pVDZ-DK type were employed ($spd$ functions for Cu and $sp$ functions for O and
Li) \cite{Balabanov_2005,Dunning_1989,Prascher_2010}.

The MCSCF calculations were carried out by `freezing' all orbitals centered at O and metal sites
beyond the reference CuO$_4$ plaquette as in the initial Hartree-Fock computation.
To separate the transition-metal $3d$ and O $2p$ valence orbitals into different groups, i.e.,
central-plaquette and adjacent-ion orbitals, we used the Pipek-Mezey localization module
\cite{localization_PM} available in {\sc molpro}.
Separate (state-averaged) MCSCF optimizations were performed for the Cu $2p^63d^9$ and $2p^53d^{10}$
orbital configurations, with the central-plaquette Cu $3d$ and Cu $2p$, $3d$ orbitals, respectively,
as active orbitals.
The subsequent MRCI treatment was carried out with the restriction of having at most five Cu $2p$
electrons for internal and semi-internal substitutions in the case of the Cu $2p^53d^{10}$
states.
Only electrons of ions on the central plaquette were correlated by MRCI, i.e., Cu $2s$, $2p$,
$3s$, $3p$, $3d$ and O $2s$, $2p$.
Spin-orbit (SO) effects were accounted for by diagonalizing the Breit-Pauli SO matrix in the
basis of correlated scalar-relativistic states.

\section*{References}


\begin{thebibliography}{10}
\expandafter\ifx\csname url\endcsname\relax
  \def\url#1{{\tt #1}}\fi
\expandafter\ifx\csname urlprefix\endcsname\relax\def\urlprefix{URL }\fi
\providecommand{\eprint}[2][]{\url{#2}}

\bibitem{book_fulde_12}
Fulde P 2012 {\em Correlated Electrons in Quantum Matter\/} (World Scientific,
  Singapore)

\bibitem{book_veenendal_15}
van Veenendal M 2015 {\em Theory of Inelastic Scattering and Absorption of
  X-rays\/} (Cambridge Univ.~Press, Cambridge)

\bibitem{XA_noci_yang96}
Yang L, Agren H, Carravetta V and Pettersson L~G~M 1996 {\em Phys. Scripta\/}
  {\bf 54} 614

\bibitem{XA_raspt2_pinjari14}
Pinjari R~V, Delcey M~G, Guo M, Odelius M and Lundberg M 2014 {\em J. Chem.
  Phys.\/} {\bf 141} 124116

\bibitem{rixs_raspt2_kunnus16}
Kunnus K, Zhang W, Delcey M~G, Pinjari R~V, Miedema P~S, Schreck S, Quevedo W,
  Schr\"o{}der H, F\"o{}hlisch A, Gaffney K~J, Lundberg M, Odelius M and Wernet
  P 2016 {\em J. Phys. Chem. B\/} {\bf 120} 7182

\bibitem{rixs_raspt2_guo16}
Guo M, Kallman E, S{\o}rensen L~K, Delcey M~G, Pinjari R~V and Lundberg M 2016
  {\em J. Phys. Chem. A\/} {\bf 120} 5848

\bibitem{RIXS_RevModPhys_11}
Ament L~J~P, van Veenendaal M, Devereaux T~P, Hill J~P and van~den Brink J 2011
  {\em Rev. Mod. Phys.\/} {\bf 83} 705

\bibitem{rixs_mgn_2009}
Braicovich L, Ament L~J~P, Bisogni V, Forte F, Aruta C, Balestrino G, Brookes
  N~B, Luca G~M~D, Medaglia P~G, Miletto-Granozio F, Radovic M, Salluzzo M,
  van~den Brink J and Ghiringhelli G 2009 {\em Phys. Rev. Lett.\/} {\bf 102}
  167401

\bibitem{rixs_d1_2009}
Ulrich C, Ament L~J~P, Ghiringhelli G, Braicovich L, Sala M~M, Pezzotta N,
  Schmitt T, Khaliullin G, van~den Brink J, Roth H, Lorenz T and Keimer B 2009
  {\em Phys. Rev. Lett.\/} {\bf 103} 107205

\bibitem{Moretti_CFfits_2011}
Moretti~Sala M, Bisogni V, Aruta C, Balestrino G, Berger H, Brookes N~B,
  de~Luca G~M, Di~Castro D, Grioni M, Guarise M, Medaglia P~G, Miletto~Granozio
  F, Minola M, Perna P, Radovic M, Salluzzo M, Schmitt T, Zhou K~J, Braicovich
  L and Ghiringhelli G 2011 {\em New J. Phys.\/} {\bf 13} 043026

\bibitem{rixs_orbiton_2012}
Schlappa J, Wohlfeld K, Zhou K~J, Mourigal M, Haverkort M~W, Strocov V~N, Hozoi
  L, Monney C, Nishimoto S, Singh S, Revcolevschi A, Caux J~S, Patthey L,
  R\o{}nnow H~M, van~den Brink J and Schmitt T 2012 {\em Nature\/} {\bf 485} 82

\bibitem{rixs_phonons_2013}
Lee W~S, Johnston S, Moritz B, Lee J, Yi M, Zhou K, Schmitt T, Patthey L,
  Strocov V, Kudo K, Koike Y, van~den Brink J, Devereaux T~P and Shen Z~X 2013
  {\em Phys. Rev. Lett.\/} {\bf 110} 265502

\bibitem{rixs_phonons_2016}
Johnston S, Monney C, Bisogni V, Zhou K, Kraus R, Behr G, Strocov V, M\'alek J,
  Drechsler S, Geck J, Schmitt T and van~den Brink J 2016 {\em Nat. Commun.\/}
  {\bf 7} 10653

\bibitem{rixs_nano_mark_2012}
Dean M~P~M, Springell R~S, Monney C, Zhou K~J, Pereiro J, Bozovic I, Piazza
  B~D, R\o{}nnow H~M, Morenzoni E, van~den Brink J, Schmitt T and Hill J~P 2012
  {\em Nature Mat.\/} {\bf 11} 850

\bibitem{rixs_nano_lucio_2012}
Minola M, Castro D~D, Braicovich L, Brookes N~B, Innocenti D, Sala M~M, Tebano
  A, Balestrino G and Ghiringhelli G 2012 {\em Phys. Rev. B\/} {\bf 85} 235138

\bibitem{rixs_ament_2007}
Ament L~J~P, Forte F and van~den Brink J 2007 {\em Phys. Rev. B\/} {\bf 75}
  115118

\bibitem{rixs_UHB_freericks_2012}
Pakhira N, Freericks J~K and Shvaika A~M 2012 {\em Phys. Rev. B\/} {\bf 86}
  125103

\bibitem{rixs_UHB_igarashi_2013}
Igarashi J and Nagao T 2013 {\em Phys. Rev. B\/} {\bf 88} 014407

\bibitem{rixs_UHB_tohyama_2015}
Tohyama T, Tsutsui K, Mori M, Sota S and Yunoki S 2015 {\em Phys. Rev. B\/}
  {\bf 92} 014515

\bibitem{rixs_cdw_benjamin_2015}
Benjamin D, Klich I and Demler E 2015 {\em Phys. Rev. B\/} {\bf 92} 035151

\bibitem{rixs_NiO_wray_2012}
Wray L~A, Yang W, Eisaki H, Hussain Z and Chuang Y~D 2012 {\em Phys. Rev. B\/}
  {\bf 86} 195130

\bibitem{rixs_tanaka_2011}
Ikeno H, Mizoguchi T and Tanaka I 2011 {\em Phys. Rev. B\/} {\bf 83} 155107

\bibitem{rixs_haverkort_2012}
Haverkort M~W, Zwierzycki M and Andersen O~K 2012 {\em Phys. Rev. B\/} {\bf 85}
  165113

\bibitem{rixs_qc_josefsson_2012}
Josefsson I, Kunnus K, Schreck S, F{\"o}hlisch A, de~Groot F, Wernet P and
  Odelius M 2012 {\em J. Phys. Chem. Lett.\/} {\bf 3} 3565--3570

\bibitem{rixs_qc_kunnus_2013}
Kunnus K, Josefsson I, Schreck S, Quevedo W, Miedema P~S, Techert S, de~Groot
  F~M~F, Odelius M, Wernet P and F\"o{}hlisch A 2013 {\em J. Phys. Chem. B\/}
  {\bf 117} 16512--16521

\bibitem{rixs_qc_maganas_2014}
Manganas D, Kristiansen P, Duda L~C, Knop-Gericke A, De{B}eer S, Schl{\"o}gl R
  and Neese F 2014 {\em J. Phys. Chem. C\/} {\bf 118} 20163--20175

\bibitem{book_QC_00}
Helgaker T, J{\o}rgensen P and Olsen J 2000 {\em Molecular Electronic-Structure
  Theory\/} (Wiley, Chichester)

\bibitem{noci_broer_81}
Broer R and Nieuwpoort W~C 1981 {\em Chem. Phys.\/} {\bf 54} 291

\bibitem{SI_malmqvist_86}
Malmqvist P~{\AA} 1986 {\em Int. J. Quantum Chem.\/} {\bf 30} 479

\bibitem{SI_mitrushchenkov_07}
Mitrushchenkov A~O and Werner H~J 2007 {\em Mol. Phys.\/} {\bf 105} 1239

\bibitem{DKH_1974}
Douglas M and Kroll N~M 1974 {\em Ann. Phys.\/} {\bf 82} 89--155

\bibitem{DKH_1986}
Hess B~A 1986 {\em Phys. Rev. A\/} {\bf 33} 3742--3748

\bibitem{Balabanov_2005}
Balabanov N~B and Peterson K~A 2005 {\em J. Chem. Phys.\/} {\bf 123} 064107

\bibitem{Dunning_1989}
Dunning~Jr T~H 1989 {\em J. Chem. Phys.\/} {\bf 90} 1007

\bibitem{Prascher_2010}
Prascher B~P, Woon D~E, Peterson K~A, Dunning T~H and Wilson A~K 2010 {\em
  Theor. Chem. Acc.\/} {\bf 128} 69--82

\bibitem{Molpro12}
Werner H~J, Knowles P~J, Knizia G, Manby F~R and Sch\"{u}tz M 2012 {\em Wiley
  Rev: Comp. Mol. Sci.\/} {\bf 2} 242--253

\bibitem{book_Dirac_TPQM}
Dirac P~A~M 1958 {\em The principles of quantum mechanics\/} fourth (revised)
  ed (Oxford University Press)

\bibitem{noci_degraaf_99}
Bagus P~S, Broer R, de~Graaf C and Nieuwpoort W~C 1999 {\em J. Electron
  Spectrosc. Relat. Phenom.\/} {\bf 98-99} 303

\bibitem{noci_hozoi_06}
Hozoi L, de~Vries A~H, Broer R, de~Graaf C and Bagus P~S 2006 {\em Chem.
  Phys.\/} {\bf 331} 178

\bibitem{diamond_alex_14}
Stoyanova A, Mitrushchenkov A, Hozoi L, Stoll H and Fulde P 2014 {\em Phys.
  Rev. B\/} {\bf 89} 235121

\bibitem{mgo_hozoi_07}
Hozoi L, Birkenheuer U, Fulde P, Mitrushchenkov A and Stoll H 2007 {\em Phys.
  Rev. B\/} {\bf 76} 085109

\bibitem{CuO_hozoi_07}
Hozoi L and Laad M~S 2007 {\em Phys. Rev. Lett.\/} {\bf 99} 256404

\bibitem{noci_vanoosten_96}
van Oosten A~B, Broer R and Nieuwpoort W~C 1996 {\em Chem. Phys. Lett.\/} {\bf
  257} 207

\bibitem{noci_hozoi_03}
Broer R, Hozoi L and Nieuwpoort W~C 2003 {\em Mol. Phys.\/} {\bf 101} 233

\bibitem{Wizent_2014}
Wizent N, Leps N, Behr G, Klingeler R, B\"uchner B and L\"oser W 2014 {\em J.
  Cryst. Growth\/} {\bf 401} 596--600

\bibitem{rixs_adress_2010}
Strocov V~N, Schmitt T, Flechsig U, Schmidt T, Imhof A, Chen Q, Raabe J,
  Betemps R, Zimoch D, Krempasky J, Wang X, Grioni M, Piazzalunga A and Patthey
  L 2010 {\em J. Synchrotron Radiat.\/} {\bf 17} 631

\bibitem{rixs_ZR_2013}
Monney C, Bisogni V, Zhou K~J, Kraus R, Strocov V~N, Behr G, M\'alek J, Kuzian
  R, Drechsler S~L, Johnston S, Revcolevschi A, B\"u{}chner B, R\o{}nnow H~M,
  van~den Brink J, Geck J and Schmitt T 2013 {\em Phys. Rev. Lett.\/} {\bf 110}
  087403

\bibitem{Learmonth_2007}
Learmonth T, McGuinness C, Glans P~A, Downes J~E, Schmitt T, Duda L~C, Guo J~H,
  Chou F~C and Smith K~E 2007 {\em Europhys. Lett.\/} {\bf 79} 47012

\bibitem{rixs_qc_carniato_2009}
Carniato S, Guillemin R, Stolte W~C, Journel L, Ta{\"i}eb R, Lindle D~W and
  Simon M 2009 {\em Phys. Rev. A\/} {\bf 80} 032513

\bibitem{rixs_qc_kavcic_2010}
Kav{\v{c}}i{\v{c}} M, {\v{Z}}itnik M, Bu{\v{c}}ar K, Miheli{\v{c}} A, Carniato
  S, Journel L, Guillemin R and Simon M 2010 {\em Phys. Rev. Lett.\/} {\bf 105}
  113004

\bibitem{QPs_hirsch_2014}
Hirsch J~E 2014 {\em Phys. Rev. B\/} {\bf 90} 104501

\bibitem{QPs_reining_2015}
Zhou J~S, Kas J~J, Sponza L, Reshetnyak I, Guzzo M, Giorgetti C, Gatti M,
  Sottile F, Rehr J~J and Reining L 2015 {\em J. Chem. Phys.\/} {\bf 143}
  184109

\bibitem{xps_rehr_2015}
Kas J~J, Vila F~D, Rehr J~J and Chambers S~A 2015 {\em Phys. Rev. B\/} {\bf 91}
  121112

\bibitem{Tischler_2003}
Tischler J~Z, Larson B~C, Zschack P, Fleszar A and Eguiluz A~G 2003 {\em Phys.
  Status Solidi B\/} {\bf 237} 280--288

\bibitem{Gurtubay_2005}
Gurtubay I~G, Pitarke J~M, Ku W, Eguiluz A~G, Larson B~C, Tischler J, Zschack P
  and Finkelstein K~D 2005 {\em Phys. Rev. B\/} {\bf 72}

\bibitem{qc_NNs_degraaf_99}
de~Graaf C, Sousa C and Broer R 1999 {\em J. Mol. Struct. (Theochem)\/} {\bf
  458} 53

\bibitem{Na2V2O5_hozoi_02}
Hozoi L, de~Vries A~H, van Oosten A~B, Broer R, Cabrero J and de~Graaf C 2002
  {\em Phys. Rev. Lett.\/} {\bf 89} 076407

\bibitem{CuO2_dd_hozoi_11}
Hozoi L, Siurakshina L, Fulde P and van~den Brink J 2011 {\em Sci. Rep.\/} {\bf
  1} 65

\bibitem{NiO_domingo_12}
Domingo A, Rodr\'{i}guez-Fortea A, Swart M, de~Graaf C and Broer R 2012 {\em
  Phys. Rev. B\/} {\bf 85} 155143

\bibitem{Os227_bogdanov_12}
Bogdanov N~A, Maurice R, Rousochatzakis I, van~den Brink J and Hozoi L 2013
  {\em Phys. Rev. Lett.\/} {\bf 110} 127206

\bibitem{SIA_Fe_maurice_2013}
Maurice R, Verma P, Zadrozny J~M, Luo S, Borycz J, Long J~R, Truhlar D~G and
  Gagliardi L 2013 {\em Inorg. Chem.\/} {\bf 52} 9379

\bibitem{localization_PM}
Pipek J and Mezey P~G 1989 {\em J. Chem. Phys.\/} {\bf 90} 9

\end{thebibliography}

\providecommand{\newblock}{}

\end{document}